\documentclass[aps,pre,twocolumn,groupedaddress]{revtex4-2}
\usepackage{tikz}
\usepackage{footnote}

\usepackage{caption}
\usepackage{amsmath,amssymb}
\usepackage{cleveref}

\usepackage[noend]{algpseudocode} 
\usepackage{algorithm}

\usepackage{bbm}

\newcommand{\ub}{\mathcal{U}}
\newcommand{\lb}{\hat{\mathcal{U}}}
\crefname{figure}{figure}{figures}
\Crefname{figure}{Figure}{Figures}
\crefname{algorithm}{Algorithm}{figures}
\Crefname{algorithm}{Algorithm}{Figures}
\crefname{corollary}{Corollary}{corollary}
\Crefname{corollary}{Corollary}{corollary}
\crefname{section}{Section}{section}
\Crefname{section}{Section}{section}
\newcommand{\mc}[1]{\mathcal{#1}}
\DeclareRobustCommand{\coh}{\mathcal{V}}

\newcounter{comments}

\begin{document}


\title{Upper bound for the stability of Boolean networks}


\author{Venkata Sai Narayana Bavisetty}
\email{narayana.venkata@medicine.ufl.edu}
\affiliation{University of Florida}

\author{Matthew Wheeler}
\email{matthew.wheeler@medicine.ufl.edu}
\affiliation{University of Florida}

\author{Reinhard Laubenbacher}
\email{reinhard.laubenbacher@medicine.ufl.edu}
\affiliation{University of Florida}

\author{Claus Kadelka}
\email{ckadelka@iastate.edu}
\affiliation{Iowa State University}


\date{\today}

\begin{abstract}
Boolean networks, inspired by gene regulatory networks, were developed to understand the complex behaviors observed in biological systems, with network attractors corresponding to biological phenotypes or cell types. In this article, we present a proof for a conjecture by Williadsen, Triesch and Wiles about upper bounds for the stability of basins of attraction in Boolean networks.
We further extend this result from a single basin of attraction to the entire network. 
Specifically, we demonstrate that the asymptotic upper bound for the robustness and the basin entropy of a Boolean network are negatively linearly related.

\end{abstract}


\maketitle

\section{Introduction}
Boolean networks were first introduced by Kauffman as models for gene regulatory networks \cite{kauffmanfirst}. 
Over the past decades, these models have gained popularity because of their simplicity and ability to qualitatively capture the complex behaviors of biological systems.
For example, Boolean network models have been applied effectively to understand intricate biological mechanisms, including the cell cycle in fission yeast \cite{davidich2008boolean}, EGFR signaling \cite{EGFR}, and CD4+ T cell differentiation \cite{tcell}. Several repositories contain hundreds of biological Boolean network models~\cite{helikar2012cell,kadelka2024meta,biodivine}. 
The long-term behavior of these models reveals important information about the phenotypes of the underlying biological system and can thus help identify potential interventions to shift the system from a disease to a healthy phenotype.


Systematic investigations of these biological models revealed that they are surprisingly robust. 
That is, they are resilient to perturbations and tend to reach the same steady state, i.e., small perturbations mostly fail to perturb the phenotype of a  biological system. 
An explanation of this robust behavior was first given by Kauffman, who empirically showed that the connectivity of a Boolean network, which is typically very low for biological systems, is inversely proportional to its robustness~\cite{kauffmanfirst}.
Furthermore, he showed that the robustness of random Boolean networks undergoes a phase transition at an average connectivity of two: 
Boolean networks with an average degree greater than two exhibit chaotic dynamics, while the dynamics are ordered if the average degree is below two. 
Derrida used probabilistic techniques to provide a theoretical explanation for the above bifurcation and introduced what has become known as the Derrida plot or Derrida values~\cite{derrida}, which describe the average size of a small perturbation after one update of the Boolean network.  
Numerous subsequent studies focused on the relationship between certain properties of a Boolean network and its robustness~\cite{willadsenwiles,kadelka2017influence,daniels2018criticality,manicka2022effective,park2023models,kadelka2024meta}.

Boolean networks with $N$ nodes can be thought of as dynamical systems on the $N$-dimensional hypercube graph $Q_N$ (see \cite{hypercuberef} for a survey about the hypercube graph). The vertices correspond to the $2^N$ states of the network and there is an edge from vertex $\mathbf x$ to vertex $\mathbf y$ if the network can transition from $\mathbf x$ to $\mathbf y$.
A natural question within this framework is how parameters such as connectivity affect the geometry of the network attractors (and their basins of attraction) on the hypercube.
Willadsen, Triesch, and Wiles attempted to address this question with a systematic study of the hypercube geometry in Boolean networks~\cite{willadsenwiles}. 
They introduced a measure of stability called coherence, which describes the likelihood that two neighboring states  eventually transition to the same network attractor.
Using this measure, they showed that the basins of attraction of a highly connected Boolean network are basically randomly distributed in the hypercube.
On the contrary, the basins of a Boolean network with low-degree tend to be quite structured and have higher coherence.
Furthermore, they conjectured based on empirical evidence that the coherence of a basin $B$ is asymptotically bounded by $\log_2(|B|)/N$ (\cite[Eq.~(6)]{willadsenwiles}).

This paper is organized as follows.
In Section \ref{sec:booleannetworks}, we introduce Boolean networks, the main objects of our study.
In Section \ref{sec:coherence}, we derive formulae for coherence, the measure of Boolean network robustness introduced by Willadsen, Triesch and Wiles in \cite{willadsenwiles}.
In Section \ref{sec:basincoherenceupperbound}, we use combinatorial methods developed independently in graph and coding theory (see e.g. \cite{harpercodes,harperfullproof,hsedges}) to prove a key conjecture by Willadsen, Triesch and Wiles. 
In Section \ref{sec:networkcoherenceub}, we generalize this result from a single basin of attraction to the entire network by deriving a tight asymptotic upper bound for the coherence of a Boolean network.
Furthermore, we show that the asymptotic upper bound is a linear function of the entropy of the network. 

\section{Boolean Networks}\label{sec:booleannetworks}
A Boolean network consists of $N$ nodes $x_1, \ldots, x_N$, where each node can exist in one of two states: true or false, frequently represented by 1 and 0, respectively.
Furthermore, each node $x_i$ possesses an update rule $$f_i:\{0,1\}^N \to \{0,1\},$$
which describes the state of $x_i$ at the next time step given the current state of the network. 
In this article, we only consider synchronous Boolean networks, in which all nodes are updated at the same time.
More precisely, a (synchronous) Boolean network is given by a function 
    \begin{align}
        F: \{0,1\}^N &\longrightarrow \{0,1\}^N\\
        x & \longmapsto (f_1(x),\ldots,f_n(x)).
    \end{align}
The iterations of the function $F$ describe the discrete time evolution of the state of the network.

\begin{figure}[h]
    \centering
    \begin{tikzpicture}

    \node (00) at (0,0) {(0,0)};
    \node (01) at (2,0) {(0,1)};
    \node (11) at (2,2) {(1,1)};
    \node (10) at (0,2) {(1,0)};

    \draw[dashed] (00) -- (01) -- (11) -- (10) -- (00);

    \draw[->] (00) to[out=135,in=225,looseness=8] (00);
    
    \draw[->] (11) to[out=45,in=135,looseness=8] (11);

    \draw[->] (01) to[bend left=30] (10);
    
    \draw[->] (10) to[bend left=30] (01);

\end{tikzpicture}
    \caption{The state space for the synchronous 2-node Boolean network $f(x,y)=(y,x)$. The states $(0,0)$ and $(1,1)$ are fixed points, while the states $(0,1)$ and $(1,0)$ form a limit cycle of length $2$.}
    \label{fig:exampleBN}
\end{figure}
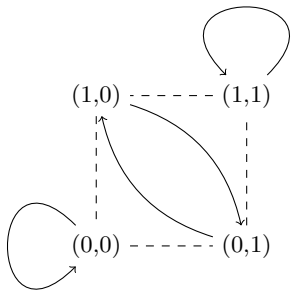

The dynamics of a synchronous Boolean network are deterministic and fully described by the \emph{state space} (also known as state transition graph), which contains $2^N$ vertices of the form $\mathbf x = (x_1,\dots,x_N) \in \{0,1\}^N$ and an edge from $\mathbf x$ to $\mathbf y$ if $F(\mathbf x) = \mathbf y$. 
The vertices of the state space coincide with the vertices of the $N$-dimensional Boolean hypercube. We are primarily interested in the long-term behavior of Boolean networks. Due to their finite size and deterministic updates, synchronous Boolean networks eventually exhibit periodic behavior. The states in a periodic orbit form a network attractor. Periodic orbits of length $1$ are called steady states or fixed points, while the states in a periodic orbit of length $m$ form an $m$-cycle. In a meaningful biological Boolean network, the attractors correspond to different phenotypes or cell types. \Cref{fig:exampleBN} shows an example of a state space of a 2-node synchronous Boolean network with different types of attractors.

Due to the determinism of the synchronous update scheme, each state can only transition into one network attractor. All states that eventually transition into this attractor form its \emph{basin of attraction}, or for short just \emph{basin}. 
That is, a synchronous $N$-node Boolean network possesses a function $$A_F: \{0,1\}^N \longrightarrow \mathcal A(F),$$ which assigns to each state its attractor, where $\mathcal A(F)$ denotes the set of all network attractors. Elements in $\mathcal A(F)$ are steady states or limit cycles.
With this, we can define the basin of any state as the set of all states that transition to the same attractor,
$$B(\mathbf x) = \{\mathbf y\in \{0,1\}^N \ \Big|\ A_F(\mathbf x) = A_F(\mathbf y)  \}.$$
This definition ensures that $B(\mathbf{x}) = B(\mathbf{y})$ whenever $\mathbf{x}$ and $\mathbf{y}$ eventually transition to the same attractor. 
Consequently, if $\mc{C} \in \mathcal{A}(F)$ is an attractor, we define the basin of attractor $\mc{C}$, $B(\mc{C})$, as the set $B(\mathbf{x})$ where $\mathbf{x}$ is any state in the attractor $\mc{C}$. 
We thus have
$$\sum_{\mc{C}\in \mathcal A(F)}|B(\mc{C})| = 2^N.$$


\section{Coherence}\label{sec:coherence}

In order to analyze the stability of a Boolean network, we use a measure called coherence, introduced by Willadsen, Triesch and Wiles~\cite{willadsenwiles}, also known as phenotypical robustness~\cite{kadelka2023modularity}.
Given a state $\mathbf x$ of a Boolean network $F$, we denote by $\mathbf x \oplus e_i$  the state where the $i$-th bit in $\mathbf x$ is flipped.
The coherence of state $\mathbf x$, denoted $\psi_{\mathbf x}$, is the fraction of neighboring states $\mathbf x \oplus e_1, \dots, \mathbf x \oplus e_n$ that are in the same basin of attraction as $\mathbf x$. That is,
\begin{equation}
\psi_{\mathbf x} = \frac{1}{N} \sum_{i = 1 }^N \mathbbm{1}[A_F(\mathbf x) = A_F(\mathbf x \oplus e_i)],\label{eq:state_coherence}
\end{equation}
where $\mathbbm{1}[\cdot]$ is the indicator function. 
The basin coherence (network coherence) is the average coherence of states in a basin (the whole network). 
That is, the coherence of basin $B$, denoted $\psi_B$, is given by
\begin{equation}
\psi_B = \frac{1}{|B|} \sum_{\mathbf x \in B} \psi_{\mathbf x}. \label{eq:basin_robustness}
\end{equation}
Similarly, the network coherence $\psi_{F}$ is given by 
\begin{equation}
\psi_{F} = \frac{1}{2^N} \sum_{\mathbf x \in \{0,1\}^N} \psi_\mathbf x. \label{eq:network_coherence}
\end{equation}

\begin{figure}[h]
    \centering  
    \begin{tikzpicture}[scale=1.8]

  \foreach \x in {0,1} {
    \foreach \y in {0,1} {
      \foreach \z in {0,1} {
        \node[draw, circle, fill=orange] (v\x\y\z) at (\x,\y,\z) {};
      }
    }
  }

  \foreach \v in {000, 001, 010, 011, 100} {
    \node[draw, circle, fill=blue] at (v\v) {};
  }

  \foreach \x in {0,1} {
    \foreach \y in {0,1} {
      \draw (v\x\y0) -- (v\x\y1);
      \draw (v\x0\y) -- (v\x1\y);
      \draw (v0\x\y) -- (v1\x\y);
    }
  }

  \foreach \x in {0,1} {
    \foreach \y in {0,1} {
      \foreach \z in {0,1} {
        \node at (\x,\y,\z) [below right] {(\x,\y,\z)};
      }
    }
  }

\end{tikzpicture}
    \caption{Example of a 3-node Boolean network with two basins of sizes $5$ (blue nodes) and $3$ (orange nodes), respectively.  The state $(0,0,0)$ is the only state with coherence $1$. The states $(0,0,1), (0,1,0), (0,1,1)$, and $(1,1,1)$ have coherence $2/3$, while the states $(1,0,0), (1,0,1)$, and $(1,1,0)$ have coherence $1/3$. The coherence of the blue basin is $2/3$, while the coherence of the orange basin is $4/9$. Therefore, the coherence of the entire network is $\frac 18 (5\cdot \frac 23  + 3\cdot \frac 49) = \frac{14}{24}$. }
    \label{fig:coherence}
\end{figure}
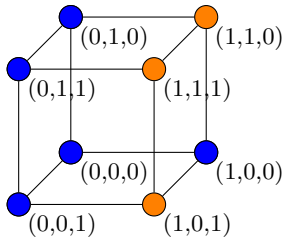

Simple algebra reveals that the coherence of the whole network is the weighted average of the coherence of the individual basins, weighted by the relative size of each basin:

\begin{equation}
\psi_{F} = \frac{1}{2^N} \sum_{B \in \mathcal{B}(F)} |B| \cdot \psi_B , \label{eq:system_coherence}
\end{equation}

where \(\mathcal{B}(F)\) is the set of disjoint basins of network $F$. \Cref{fig:coherence} exemplifies these concepts for a 3-node Boolean network.

Using combinatorial counting arguments, we now derive a formula for the coherence of a basin, which will allow us to use the results from \cite{hsedges,harperfullproof,harpercodes}. 
Following the notation in \cite{hsedges}, let $E(B,B)$ denote the set of edges of the hypercube connecting the basin $B$ with itself. 
Since the coherence of a basin is the average coherence of the states within it, we get
$$\psi_B = \frac{1}{N \cdot |B|}\sum_{\mathbf x \in B} \sum_{i=1}^N \mathbbm{1}[\mathbf x \oplus e_i \in B].$$
Each edge in $E(B,B)$ contributes exactly twice to the sum in the numerator. Thus, 
\begin{equation}\label{eq:connectiontoedgesonncube}\psi_B = \frac{2 \cdot|E(B,B)|}{N\cdot|B|}.\end{equation}

\section{Upper bound for basin coherence}\label{sec:basincoherenceupperbound}

The coherence of a basin strongly depends on the size of the basin. 
This dependence must be properly accounted for when comparing the coherence of basins of different sizes, necessitating a thorough understanding of the basin coherence.
In~\cite{willadsenwiles}, the authors revealed through computational experiments that the coherence of a basin is hardly ever smaller than the relative size of the basin. 
Furthermore, they conjectured an upper bound for the basin coherence and demonstrated that normalizing coherence by this bound facilitates a more meaningful comparison of basin stability~\cite{willadsenwilesapplied}. 
In this section, we derive an analytical expression for the upper bound of basin coherence, which proves their conjecture that the coherence of a basin $B$ is asymptotically bounded by $\log_2(|B|)/N$ (\cite[Eq.~(6)]{willadsenwiles}). 

By Eq.~\eqref{eq:connectiontoedgesonncube}, to obtain an upper bound for the coherence $\psi_B$ of a basin of size $|B| =b$, it suffices to find the maximum value of the function 
\begin{align}\label{remark:mainremark} 
\coh: \textrm{vertex subsets of }Q_N &\to \mathbb{N}\\
S\subset V(Q_N) &\mapsto |E(S,S)|, 
\end{align}
subject to the constraint $|S|=b$, where $Q_N$ denotes the $N$-dimensional hypercube graph.
 This maximization problem was studied in \cite{hsedges,harpercodes,harperfullproof}, where it naturally arises as a problem in coding theory \cite{harpercodes} and in game theory \cite{hsedges}. 
 We briefly recall previous results below.

\begin{figure}[h]
    \centering
    \begin{tikzpicture}[scale=1]







        \coordinate (A1) at (3,0,0);
        \coordinate (B1) at (4,0,0);
        \coordinate (C1) at (4,1,0);
        \coordinate (D1) at (3,1,0);
        \coordinate (E1) at (3,0,1);
        \coordinate (F1) at (4,0,1);
        \coordinate (G1) at (4,1,1);
        \coordinate (H1) at (3,1,1);

        \coordinate (A2) at (5.5,0.5,0.5);
        \coordinate (B2) at (6.5,0.5,0.5);
        \coordinate (C2) at (6.5,1.5,0.5);
        \coordinate (D2) at (5.5,1.5,0.5);
        \coordinate (E2) at (5.5,0.5,1.5);
        \coordinate (F2) at (6.5,0.5,1.5);
        \coordinate (G2) at (6.5,1.5,1.5);
        \coordinate (H2) at (5.5,1.5,1.5);



        \draw[thick] (A1) -- (B1) -- (C1) -- (D1) -- cycle;
        \draw[thick] (E1) -- (F1) -- (G1) -- (H1) -- cycle;
        \draw[thick] (A1) -- (E1);
        \draw[thick] (B1) -- (F1);
        \draw[thick] (C1) -- (G1);
        \draw[thick] (D1) -- (H1);

        \draw[thick] (A2) -- (B2) -- (C2) -- (D2) -- cycle;
        \draw[thick] (E2) -- (F2) -- (G2) -- (H2) -- cycle;
        \draw[thick] (A2) -- (E2);
        \draw[thick] (B2) -- (F2);
        \draw[thick] (C2) -- (G2);
        \draw[thick] (D2) -- (H2);

        \foreach \i/\j in {A1/A2, B1/B2, C1/C2, D1/D2, E1/E2, F1/F2, G1/G2, H1/H2}{
            \draw[thin,opacity=0.3] (\i) -- (\j);
        }

        \foreach \pt in {A1,B1,C1,D1,E1,F1,G1,H1,A2}{
            \filldraw[fill=blue,draw=black] (\pt) circle (3pt);
        }
        \foreach \pt in {F2,G2,H2}{
            \filldraw[fill=white,draw=black] (\pt) circle (3pt);
        }
        \foreach \pt in {B2,C2,D2,E2}{
            \filldraw[fill=orange,draw=black] (\pt) circle (3pt);
        }

    \end{tikzpicture}
    \caption{Example of a 4-node Boolean network with three basins of size $9$ (blue), $4$ (orange) and $3$ (white), respectively. The basins in blue and white are both Harper arrays, whereas the basin in orange is not a Harper array.}
    \label{fig:harpercube}
\end{figure}
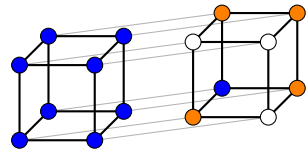

In \cite{harpercodes}, the authors introduce Harper arrays and show that all solutions to the above maximization problem must be Harper arrays.
In the following, a $k$-subcube is a subset of $Q_N$ of size $2^k$, where $N-k$ coordinates are fixed. Furthermore,  a shadow of a $k$-subcube is another $k$-subcube obtained by complementing exactly one of the $N-k$ fixed coordinates.
A Harper array $S$ of size $b$ consists of a sequence of subcubes of dimensions $i_1, i_2, \ldots, i_s$, where $i_1 > i_2 > \ldots > i_s$  and $b = 2^{i_1} + 2^{i_2} + \ldots + 2^{i_s}$, such that each subcube of dimension $i_j$ is in the shadow of every subcube of dimension larger than $i_j$. As an example, consider the blue and the white basins in \Cref{fig:harpercube}, which are Harper arrays of size 9 and 3, respectively. On the contrary, the orange basin of size 4 is not a Harper array.

We are now ready to give an analytical proof of the conjecture in \cite{willadsenwiles}.
In \protect{\cite{harpercodes},\cite{harperfullproof},\cite[Theorem 1.4]{hsedges}}, the authors show that a subset $S$ of hypercube $Q_N$ is a Harper array if and only if it maximizes the function $\coh$.
Therefore, for a basin $B$, we have the inequality:
\begin{equation}
    \psi_B \leq \frac{2\cdot|E(S,S)|}{N \cdot |S|}, 
\end{equation}
where $S$ is a Harper array with the same cardinality as $B$.
For a Harper array $S$, the cardinality $|E(S,S)|$ is given by  $\ub(|S|)$, where $\ub:\mathbb{N} \to \mathbb{N} $  can be computed recursively as follows:
    \begin{align}\label{definition:upperbound}
    \ub(1) &= 0 \nonumber \\
    \ub(n) &= \ub(n{-}1) + \quad
             \substack{\text{\normalsize number of 1s in the binary}\\\text{\normalsize representation of $n{-}1$}}.
\end{align}

The first eight entries of this sequence are $0, 1, 2, 4, 5, 7, 9, 12$ (see~\cite{oeisA000788}), and we obtain the upper bound
\begin{equation}\label{eq:coherenceub}
    \psi_B \leq \frac{2\cdot \ub(|B|)}{N \cdot |B|},
\end{equation}
 where equality is achieved if and only if the basin $B$ is a Harper array.
In \cite{asymptotics, Tang}, the authors show that, for any natural number $n$, we have
\begin{equation}\label{eq:asymptotic}\ub(n) = \frac{n \log_2(n)}{2} + O(n).\end{equation}
Substituting the term $\ub(|B|)$ in Eq.~\eqref{eq:coherenceub} with its asymptotic estimate $\frac{|B| \log_2(|B|)}{2} + O(|B|)$, we get that $\frac{\log_2(|B|)}{N}$ is an asymptotic upper bound for the basin coherence, i.e.,
    \begin{equation}
       \psi_B \leq \frac{\log_2(|B|)}{N} + O\left(\frac{1}{N}\right).
    \end{equation}

While this limits the coherence of small basins in small Boolean networks, small basins in large networks may still exhibit a high level of coherence. For instance, when $N = 100$, a coherent basin comprising just $10\%$ of the state space can attain a coherence value of up to approximately $0.97$. Interestingly, smaller basins in sparsely connected Boolean networks often display significantly higher coherence than would be expected based on their size~\cite{willadsenwiles}. 
This includes most biological network models, most of which possess an average degree of below 3~\cite{kadelka2024meta}.

    \section{Upper bound for the coherence of a Boolean network}\label{sec:networkcoherenceub}
    
    In the previous section, we discussed the problem of finding the maximum coherence of a single basin of attraction. 
    In this section, we extend these results by deriving a tight asymptotic upper bound for the coherence of an entire Boolean network $F$ on $N$ nodes with $k$ basins of size $n_1, \ldots, n_k$ respectively, where $n_1+\ldots+n_k = 2^N$.
   
    Given the connection between basin and network coherence (Eq.~\ref{eq:system_coherence}), a natural way to find an upper bound for the coherence $\psi_{F}$ of such a network is to use the upper bound for the coherence of each basin (Eq.~\ref{eq:coherenceub}). 
    This yields the following upper bound for the network coherence 
    \begin{align}\label{eq:upperbound}
       \psi_{F} &\leq  \frac 1{N \cdot 2^{N-1}}\sum \limits_{i=1}^k \ub(n_i), 
    \end{align}
    where  $\ub(n)$ is the recursive function defined in Eq.~\eqref{definition:upperbound}.
    Furthermore, this upper bound is tight if and only if there exists a Boolean network with basin sizes $n_1,\ldots,n_k$, such that each basin is a Harper array. 

    If $F$ is an $N$-node Boolean network with $k$ basins of sizes $n_1,\ldots,n_k$, which are all Harper arrays,
    then, using Eq.~\eqref{eq:asymptotic}, the coherence of $F$ can be expressed as
\begin{align*}\label{eq:coherenceharpernet}
        \psi_{F} &= \frac{1}{N2^{N-1}}\sum_{i=1}^{k} \ub(n_i)\\[1ex]
          &= \sum_{i=1}^k \frac{n_i}{2^N}.\frac{\log_2(n_i)}{N} + O\left(\frac{1}{N}\right).
    \end{align*}
    
The basin entropy, or for short just entropy, of a Boolean network was introduced in~\cite{krawitz2007basin} as a measure of the complexity of information that a Boolean network can store. Like the classical Shannon entropy at the foundation of information theory~\cite{shannon1948mathematical}, basin entropy is defined as
\begin{equation}\label{eq:entropy} \mathcal{E}_F := -\sum_{i=1}^k b_i \log_2\left(b_i\right), \end{equation}
where $b_i = n_i/2^N$ describes the normalized size of each basin of attraction. A Boolean network with a single attractor has basin entropy 0, while any network with multiple attractors has positive entropy. Networks with many attractors of similar basin size have particularly high entropy. 

To derive the relationship between network coherence and entropy, we rearrange the expression for the entropy in Eq.~\eqref{eq:entropy}:
\begin{align}\label{eq:coherenceentropy}
    \mc{E}_{F} &= N -\frac{1}{2^N} \left(\sum_{i=1}^k n_i \log_2(n_i)\right)\\
    &= N-N\cdot\psi_{F} + O(1).
\end{align}

    Therefore, the coherence and entropy of a Boolean network, in which every basin is a Harper array, are asymptotically negatively linearly related:
    \begin{align}\label{eq:entropycoherence}
    \psi_{F} = 1 - \frac{\mc{E}_{F}}N + O\left(\frac 1N\right).
\end{align}

    However, it is not always possible to construct such a network. For any $N\geq 5$, there exist integer partitions $n_1,\dots,n_k$ of $2^N$ such that it is impossible for all basins of sizes $n_1,\dots,n_k$ to be Harper arrays.
    For example, a 5-node Boolean network with basins of size 15, 13, and 4 such that all the basins are Harper arrays does not exist. This is illustrated in  \Cref{fig:counterexample}. 
    The basin of size $15$ (indicated in blue) must occupy one half of the hypercube, i.e., it must be contained in a 4-cube. 
    This leaves a single vertex in that 4-cube unassigned (circled in red), which must then belong to either of the other two basins.
However, this vertex cannot be part of a Harper array of size $4$, as there exists no empty 2-cube containing it. 
Similarly, it cannot be included in a Harper array of size $13$, since any 4-dimensional subcube containing this vertex has only 9 or fewer available positions remaining (i.e., after placing the basin of size 15). These considerations show why it impossible for a 5-node Boolean network to have basins of sizes 15, 13, and 4,  which are all Harper arrays.

\newcommand{\offsetA}{3}  
\newcommand{\offsetB}{2}
\newcommand{\offsetC}{0.5}

    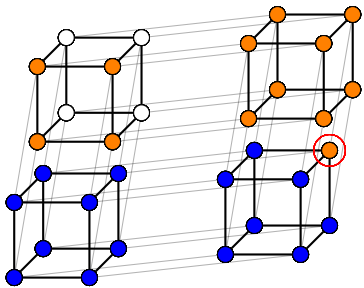
\begin{figure}
        \centering
        \begin{tikzpicture}
\foreach \name/\x/\y/\z in {
    A1/0/0/0, B1/1/0/0, C1/1/1/0, D1/0/1/0,
    E1/0/0/1, F1/1/0/1, G1/1/1/1, H1/0/1/1
}{
    \coordinate (\name) at (\x,\y,\z);
}

\foreach \name/\x/\y/\z in {
    A2/\offsetA/0.5/0.5, B2/\offsetA+1/0.5/0.5, C2/\offsetA+1/1.5/0.5, D2/\offsetA/1.5/0.5,
    E2/\offsetA/0.5/1.5, F2/\offsetA+1/0.5/1.5, G2/\offsetA+1/1.5/1.5, H2/\offsetA/1.5/1.5
}{
    \coordinate (\name) at (\x,\y,\z);
}

\foreach \name/\x/\y/\z in {
    A3/0+\offsetC/\offsetB/0.5, B3/1+\offsetC/\offsetB/0.5, C3/1+\offsetC/\offsetB+1/0.5, D3/0+\offsetC/\offsetB+1/0.5,
    E3/0+\offsetC/\offsetB/1.5, F3/1+\offsetC/\offsetB/1.5, G3/1+\offsetC/\offsetB+1/1.5, H3/0+\offsetC/\offsetB+1/1.5
}{
    \coordinate (\name) at (\x,\y,\z);
}

\foreach \name/\x/\y/\z in {
    A4/\offsetA+\offsetC/\offsetB+0.5/1, B4/\offsetA+1+\offsetC/\offsetB+0.5/1, C4/\offsetA+1+\offsetC/\offsetB+1.5/1, D4/\offsetA+\offsetC/\offsetB+1.5/1,
    E4/\offsetA+\offsetC/\offsetB+0.5/2, F4/\offsetA+1+\offsetC/\offsetB+0.5/2, G4/\offsetA+1+\offsetC/\offsetB+1.5/2, H4/\offsetA+\offsetC/\offsetB+1.5/2
}{
    \coordinate (\name) at (\x,\y,\z);
}

\foreach \set in {1,2,3,4}{
    \foreach \a/\b/\c/\d in {A/B/C/D, E/F/G/H}{
        \draw[thick] (\a\set) -- (\b\set) -- (\c\set) -- (\d\set) -- cycle;
    }
    \foreach \a/\e in {A/E, B/F, C/G, D/H}{
        \draw[thick] (\a\set) -- (\e\set);
    }
}

\foreach \i/\j in {A1/A2, B1/B2, C1/C2, D1/D2, E1/E2, F1/F2, G1/G2, H1/H2,
                   A3/A4, B3/B4, C3/C4, D3/D4, E3/E4, F3/F4, G3/G4, H3/H4}{
    \draw[thin,opacity=0.3] (\i) -- (\j);
}

\foreach \i/\j in {A1/A3, B1/B3, C1/C3, D1/D3, E1/E3, F1/F3, G1/G3, H1/H3,
                   A2/A4, B2/B4, C2/C4, D2/D4, E2/E4, F2/F4, G2/G4, H2/H4}{
    \draw[thin,opacity=0.3] (\i) -- (\j);

    \foreach \pt in {A1,B1,C1,D1,E1,F1,G1,H1,A2,B2,D2, E2,H2,F2,G2}{
            \filldraw[fill=blue,draw=black] (\pt) circle (3pt);
        }

    \foreach \pt in {C2}{
    \filldraw[fill = orange, draw = black] (\pt) circle (3pt);
    \draw[color=red] (\pt) circle (6 pt);
    }

    \foreach \pt in {E3,F3,G3,H3,A4,B4,C4,D4, E4,H4,F4,G4}{
            \filldraw[fill=orange,draw=black] (\pt) circle (3pt);
        }

    \foreach \pt in {A3,B3,C3,D3}{
            \filldraw[fill=white,draw=black] (\pt) circle (3pt);
        }
}     \end{tikzpicture}
        \caption{Illustration why three basins of sizes $15$ (blue), $13$ (orange), and $4$ (white) within a 5-dimensional hypercube cannot all be Harper arrays. The basin of size $15$ forms a Harper array, which leaves the vertex circled in red unassigned. This vertex must be included in either the basin of size $13$ or $4$. In this configuration, it is assigned to the basin of size $13$; however, this assignment causes the orange basin to \emph{not} form a Harper array. If instead assigned to the basin of size $4$, this basin could not be a Harper array.}
        \label{fig:counterexample}
    \end{figure}

    This shows that the upper bound in Eq.~\eqref{eq:upperbound} cannot always be achieved. 
    However in Appendix~\ref{sec:constructcoherent}, we show that it is always asymptotically achieved: given basin sizes $n_1,\ldots,n_k$, there exists a network $G$ with coherence $\psi_{G}$, such that  
    
    

   \begin{equation}\label{equation:lowerbound}
   \frac{1}{N\cdot 2^{N-1}}\sum \limits_{i=1}^k \lb(n_i) \leq   \psi_{G} \leq \frac{1}{N\cdot 2^{N-1}}\sum \limits_{i=1}^k \ub(n_i),
    \end{equation} where $\lb(n)$ is defined in terms of  $\ub(n)$  as 
    \begin{equation}\label{definition:lowerbound}
\lb(n) := \sum_{j=1}^{s} \ub(2^{i_j}), \quad \text{for } n = 2^{i_1} + \ldots + 2^{i_s},  
\end{equation}
where  $i_1 > i_2 > \dots > i_s \geq 0$.
Even though there are basins in network $G$ that are not Harper arrays, 
 each basin of size $n= 2^{i_1} + \ldots + 2^{i_s}$ consists of disjoint subcubes of respective dimensions $i_1,\ldots,i_s$.
 Note that, since the basins of $G$ are not necessarily Harper arrays, there may not be the additional edges between the subcubes that emerge in a Harper array due to the embedding of smaller cubes within the shadow of larger ones.
 
    However, we show in Appendix~\ref{sec:technicallemma} that $\ub(n)$ and $\lb(n)$ have the same asymptotic behavior. That is, 
    \begin{equation}\label{eq:sameasymptote}
        \lb(n) \sim \ub(n) \sim  \frac{n \log_2(n)}{2}. 
    \end{equation}
Combining Eqs.~\eqref{equation:lowerbound}~and~\eqref{eq:sameasymptote}, we get that the upper bound for coherence is asymptotically achieved for network $G$:
    \begin{equation}
        \psi_{G} \sim \sum_{i=1}^{k} \frac{n_i}{2^N}\frac{\log_2(n_i)}{N},
    \end{equation}
    even though its basin size distribution may prevent all basins from being arranged as Harper arrays. Thus, the upper bound is asymptotically tight, irrespective of the specific basin sizes of a Boolean network. That is, we have \begin{equation}\label{eq:mprline}
           \psi_{F} \leq  1- \frac{\mc{E}_{F}}{N} + O\left(\frac{1}{N}\right).
       \end{equation}

\begin{figure}
        \centering
        \includegraphics[width=1\linewidth]{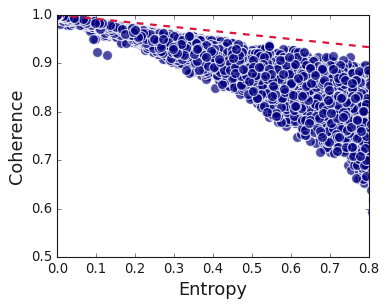}
        \caption{Coherence versus entropy for 50,000 random 12-node Boolean networks with a constant in-degree of $2$. The dashed red line represents the asymptotic upper bound derived in Eq.~\eqref{eq:mprline_without_bigO}.To emphasize the low entropy regime, where the difference between maximally observed coherence and the upper bound is smaller and potential violations of the bound may occur, data corresponding to networks with entropy greater than 0.8 have been excluded from the analysis.}
    \label{fig:experimentaldata}
    \end{figure}

  This implies that, in the limit $N \to \infty$, the maximal coherence of network is exactly negatively linearly
  related to its entropy: 
  \begin{equation}\label{eq:mprline_without_bigO}
  \psi_F \leq 1 - \mathcal{E}_F/N.\end{equation} 
  To test how well this bound works for small Boolean networks, we generated 50,000 random 12-node Boolean networks with a fixed in-degree of 2. The coherence of all these networks was below the upper bound $1-\mathcal{E}_F/{12}$~(\Cref{fig:experimentaldata}).

  
\section{Discussion}

Network and basin coherence are two interrelated metrics that quantify the stability of a Boolean network and its basins. 
Through analytical arguments, we showed that the basin coherence is in the limit logarithmically bounded by the relative size of the basin.
Moreover, we proved that the overall coherence of a Boolean network has an upper bound that decreases linearly with increasing network entropy. These theoretical results hold only in the limit. However, we showed through computational experiments that the bounds appear to hold even for much smaller Boolean networks.

Our results are particularly relevant for the analysis of the stability of biological networks.
In particular, our findings clarify the link between the stability of a network and the structure of its state space.
To illustrate this, consider two biological networks of identical size, possessing the same number of attractors with identical basin sizes, and consequently exhibiting equal entropy.
If one network possesses a lower coherence than the other, this implies that its dynamics (i.e., its state space) is more chaotic and disorganized.
In contrast, the network with higher coherence exhibits basins that more closely resemble structured configurations such as Harper arrays, that is, the states in each basin cluster together to form subcubes. 
However, if two networks do not have the same entropy, then a normalized measure of coherence would enable a better comparison of their stability. 
Our result on a tight upper bound for the network coherence provides the first step towards such a normalized measure. 
Future work towards this goal should aim to reveal how the basin sizes and other relevant network characteristics such as network size or connectedness determine the entire distribution of network coherence, not only the maximum.

Organisms typically evolve to maximize phenotypic complexity while maintaining high stability.
Therefore, we expect biological networks to exhibit coherence values approaching the theoretical upper bound. 
However, because of the logarithmic nature of this bound, even small basins of attraction can display disproportionately high coherence relative to their size, and therefore correspond to stable biological phenotypes. 


Finally, all our analysis was performed exclusively within the framework of synchronous networks. 
However, biological systems often operate on multiple time scales and exhibit asynchronous dynamics. 
Future work should therefore explore whether the observed relationship between entropy and coherence persists in the asynchronous setting.

\begin{acknowledgments}
VSNB  thanks Maria Siskaki for useful comments and suggestions. CK was partially supported by travel grant 712537 from the Simons Foundation and award 2424632 by the National Science Foundation. The research of RL was partially supported by the grants NSF DMS-2424635, NIH R01 AI135128, 
and NIH R01 HL169974-01. MW was partially supported by NSF DMS-2424635.
\end{acknowledgments}

\appendix
\section{Technical Lemma}\label{sec:technicallemma}
In this section, we show that $\ub(n)$ and $\lb(n)$ have the same asymptotic behavior, where $\ub(n)$ and $\lb(n)$ are defined as in Eqs.~\eqref{definition:upperbound} and \eqref{definition:lowerbound}, respectively.

        Let $\alpha(n) = \ub(n) - \lb(n)$. 
        First, we observe that $\alpha(n)$ satisfies the formula
        \begin{equation}\label{eq:recursivealpha}
        \alpha(n) = n - 2^{\lfloor\log_2 n\rfloor} + \alpha(n - 2^{\lfloor\log_2 n\rfloor}).
        \end{equation}
        This follows by noting that 
        $$\ub(n) = \ub(2^{\lfloor\log_2 n\rfloor}) + \ub(n - 2^{\lfloor\log_2 n\rfloor}) + n - 2^{\lfloor\log_2 n\rfloor}$$ 
        and 
        $$\lb(n) = \lb(2^{\lfloor\log_2 n\rfloor}) + \lb(n - 2^{\lfloor\log_2 n\rfloor}).$$
        
        Equipped with the recursive relation in Eq.~\eqref{eq:recursivealpha}, we use induction to prove that $\alpha(n) \leq n$ for all $n \geq 1$.
        The base case of the induction holds because $\alpha(1) = 0$.
        For the inductive step, we assume the inequality is true for all $n \leq M$. 
        We will now show that the inequality holds for $M + 1$.
        If $M + 1$ is a power of $2$, the inequality is trivial since $\alpha(M + 1) = 0$ by definition of $\lb(n)$ in Eq.~\eqref{definition:lowerbound}.
        If it is not a power of $2$, then 
       
\begin{align*}
\alpha(M + 1) &= (M + 1) - d + \alpha((M + 1) - d) \\
              &\leq 2((M + 1) - d) \\
              &\leq M + 1,
\end{align*}
where $d = 2^{\lfloor \log_2 (M + 1) \rfloor}$.
        
        The last inequality follows since $2^{\lfloor{\log_2 (M + 1)\rfloor}} \geq \frac{M + 1}{2}$. Therefore, by induction, the inequality holds for all $n \geq 1$.
        For all $n > 0$, we have $\ub(n) - \lb(n) \leq n$.
        By \cite{asymptotics,Tang}, we know that $\ub(n) = \frac{n \log_2(n)}{2} + O(n)$.
        Therefore, we have $$  \lb(n) \sim \ub(n) \sim  \frac{n \log_2(n)}{2}. $$

\section{Construction of the Coherent Network}\label{sec:constructcoherent}

In this section, we show how to construct a highly coherent network $G$ with coherence $\psi_G$, satisfying the inequality
$$\psi_{G} \geq \frac{1}{N \cdot 2^{N-1}} \sum_{i=1}^k \lb(n_i),$$
where $\lb(n)$ is the lower bound function defined in Eq.~\eqref{definition:lowerbound}. 
Importantly, our construction is more general: it does not require the specified basins to exhaust the entire state space. 
That is, given natural numbers $N, n_1, \dots, n_k$ such that $\sum_{i=1}^k n_i \leq 2^N$, we construct an N-node Boolean network $G$ with $k$ basins of sizes $n_1, \ldots, n_k$, respectively. 
These basins may represent only a subset of all possible basins in the network. 
Nevertheless, each basin $B_i$ satisfies the coherence bound
\begin{equation} 
   \psi_{B_i} \geq  \frac{2\cdot \lb(n_i)}{N \cdot n_i} \quad \textrm{ for all } 1 \leq i \leq k.
\end{equation}
\Cref{algorithm:coherent} gives a recursive construction of the desired network  $G$.
\begin{figure}
\begin{algorithm}[H]
\caption{Construct coherent network $G$}
\label{algorithm:coherent}
\begin{algorithmic}[1]
\Statex \textbf{Input:} List of basin sizes, integer $N$
\Statex \textbf{Output:} Mapping from basin to list of states
\Function{CoherentNetwork}{basin\_sizes, $N$}
    \If{$N = 1$}
        \If{$|\text{basin\_sizes}| = 1$}
            \If{\text{basin\_sizes[0]} $= 2$}
                \State \Return $\{1 \mapsto [0,1]\}$ 
            \Else
                 \State \Return $\{1 \mapsto [0]\}$
            \EndIf
        \Else
            \State \Return $\{1 \mapsto [0],\; 2 \mapsto [1]\}$
        \EndIf
    \EndIf
    \State half\_sizes $\gets$ \texttt{EmptyList}()
    \ForAll{$s$ in basin\_sizes}
        \If{$\lfloor s/2 \rfloor > 0$}
            \State append  $\lfloor s/2 \rfloor$ to half\_sizes
        \EndIf
    \EndFor
    \State arr $\gets$ \Call{CoherentNetwork}{half\_sizes, $N-1$}
    \ForAll{(basin $\mapsto$ states) in arr}
        \ForAll{$x$ in states}
            \State append $x + 2^{N-1}$ to arr[basin]
        \EndFor
    \EndFor
    \State assigned\_states $\gets$ \texttt{Set of all states in arr}
    \State available\_states $\gets$ $\{0,1,\ldots,2^N-1\}\ -$ assigned\_states
    
    \ForAll{$s$ in basin\_sizes}
        \If{$s$ is odd}
            \State new\_state $\gets$ Pop(available\_states)
            \If{$s > 1$ }
                \State append new\_state to the corresponding basin in arr
            \Else
                \State new\_key $\gets$ \texttt{len}(arr) + 1
                \State arr[new\_key] $\gets$ [new\_state]
            \EndIf    
        \EndIf
    \EndFor
    \State \Return arr
\EndFunction
\end{algorithmic}
\end{algorithm}
\end{figure}

   We prove the correctness of this construction by induction on the number of nodes $N$.
    The base case for a single node ($N=1$) is trivial.
   For the inductive step, assume the algorithm produces a network satisfying the coherence bound for all $N \leq M$. 
    We will first consider the case, where all the sizes $n_i$ are even.
    By the inductive hypothesis, there exists a network $H$ on $M$ nodes with $k$ basins $C_1, \dots, C_k$ of sizes $n_1/2, \dots, n_k/2$, such that $\frac{n_1}{2}, \ldots, \frac{n_k}{2} $, such that the basin coherences satisfies the inequality:
    $$ \psi_{C_i} \geq  \frac{4\cdot \lb(n_i/2)}{M \cdot n_i} \quad \textrm{ for all } 1 \leq i \leq k$$
       
    We embed the network $G_1$ in the hyperface $x_{M+1}=0$ of the $(M+1)$-hypercube.
   The basins of the desired network $G$ are then formed by augmenting each state $\mathbf x$ in a basin of $H$ with its reflection $\mathbf x\oplus e_{M+1}$, where $e_{M+1} = (0,\ldots,0,1)$. 
   This yields basins of sizes $n_1, \dots, n_k$ in $G$.
    
    Using the identity $$\lb\left(\frac{n_i}{2}\right)+ \lb\left( \frac{n_i}{2}\right) +  \frac{n_i}{2} = \lb(n_i),$$ we get for all $i = 1,\ldots,k$ that
$$ \psi_{B_i} \geq  \frac{2\cdot \lb(n_i)}{(M+1) \cdot n_i}.$$
Thus, the overall network coherence satisfies
    $$\psi_{G} \geq \frac{1}{(M+1)2^{M}}\sum_{i=1}^k \lb (n_i).$$
    
    Finally, if any $n_i$ is odd, we replace it with $n_i - 1$, noting that $\lb(n) = \lb(n - 1)$ for odd $n$. 
    This reduction brings us back to the even case, completing the proof.


\bibliography{refs}

\end{document}